\newif \ifconf
\conffalse
\ifconf
\documentclass{article}
\usepackage{spconf,amsmath,graphicx,hyperref}

\usepackage{amsmath}
\usepackage{amssymb}
\usepackage{amsthm}
\usepackage{xcolor}
\usepackage{dsfont}
\usepackage{enumitem}
\usepackage{comment}
\usepackage{cite}

\usepackage{tikz}
\usepackage{pgfplots}
\pgfplotsset{compat=1.18}   
\usepgfplotslibrary{groupplots}
\usepgfplotslibrary{dateplot}

\newtheorem{theorem}{Theorem}[section] 
\newtheorem{lemma}[theorem]{Lemma}
\newtheorem{corollary}[theorem]{Corollary}
\newtheorem{proposition}[theorem]{Proposition}
\newtheorem{remark}{Remark}

\theoremstyle{definition}
\newtheorem{definition}[theorem]{Definition}
\else 
\documentclass[conference,10pt]{IEEEtran}
\usepackage[dvipsnames]{xcolor}
\usepackage{enumitem}
\usepackage[utf8]{inputenc} 
\usepackage[T1]{fontenc}
\usepackage{url}
\usepackage{ifthen}
\usepackage{cite}
\usepackage[cmex10]{amsmath} 
\usepackage{adjustbox}
\usepackage{graphicx} 
\usepackage{hyperref}
\hypersetup{colorlinks=true, unicode=true, linkcolor=[rgb]{0.10,0.05,0.67}, citecolor=[rgb]{0.10,0.05,0.67}, filecolor=[rgb]{0.10,0.05,0.67}, urlcolor=[rgb]{0.10,0.05,0.67}}
\usepackage{physics}
\usepackage{setspace}
\usepackage{amssymb}
\usepackage{stfloats}
\usepackage{amsthm}
\usepackage{multirow}
\usepackage{amsfonts}
\usepackage{xcolor}
\usepackage{graphicx}
\usepackage{subcaption}
\usepackage[]{algorithmicx}
\usepackage{algpseudocode,algorithm}
\usepackage{mathtools}
\usepackage{stmaryrd}
\usepackage[mathscr]{euscript}
\usepackage{array}
\usepackage{mathrsfs}

\usepackage{soul}
\newtheorem{remark}{Remark}
\newtheorem{theorem}{Theorem}

\newtheorem{definition}{Definition}
\usepackage{authblk}
\usepackage{tabularx}
\usepackage{svg}
\usepackage{tikz}
\usepackage{bm}
\usetikzlibrary{arrows.meta}

\newtheorem{proposition}[theorem]{Proposition}

\allowdisplaybreaks
\fi

\newcommand{\off}[1]{}

\newcommand{\Kcalc}{\kappa_{calc}}
\newcommand{\KNerr}{\kappa_{n}^{err}}

\newcommand{\THcali}{\delta^{cali}}

\usepackage{comment}
\usepackage{epstopdf}

\ifconf
\title{Self-Calibrating Integrate-and-Fire Time
Encoding Machine}
\name{Maya Mekel$^{1}$, Vered Karp$^{1}$, Satish Mulleti$^{2}$, and Alejandro Cohen$^{1}$ \vspace{-0.2cm}}
\address{$^{1}$Faculty of ECE, Technion, Israel,\\
Emails: \{maya.mekel,veredlevi\}@campus.technion.ac.il, and alecohen@technion.ac.il \\
$^{2}$Department of EE, Indian Institute of Technology Bombay, India,
Email: mulleti.satish@gmail.com \vspace{-0.59cm}}
\else
\title{\huge Self-Calibrating Integrate-and-Fire Time Encoding Machine\vspace{-0.4cm}}

\author{Maya Mekel$^{1}$, Vered Karp$^{1}$, Satish Mulleti$^{2}$, and Alejandro Cohen$^{1}$\vspace{0.0cm}\\
$^{1}$Faculty of ECE, Technion, Israel,\\
Emails: \{maya.mekel,veredlevi\}@campus.technion.ac.il, and alecohen@technion.ac.il \\
$^{2}$Department of EE, Indian Institute of Technology Bombay, India,
Email: mulleti.satish@gmail.com \vspace{-0.58cm}}

\fi
\begin{document}
%
\maketitle
\begin{abstract}
In this paper, we introduce a novel self-calibrating integrate-and-fire time encoding machine (S-IF-TEM) that enables simultaneous parameter estimation and signal reconstruction during sampling, thereby effectively mitigating mismatch effects. The proposed framework is developed over a new practical IF-TEM (P-IF-TEM) setting, which extends classical models by incorporating device mismatches and imperfections that can otherwise lead to significant reconstruction errors. Unlike existing IF-TEM settings, P-IF-TEM accounts for scenarios where (i) system parameters are inaccurately known and may vary over time, (ii) the integrator discharge time after firings can vary, and (iii) the sampler may operate in its nonlinear region under large input dynamic ranges. For this practical model, we derive sampling rate bounds and reconstruction conditions that ensure perfect recovery. Analytical results establish the conditions for perfect reconstruction under self-calibration, and evaluation studies demonstrate substantial improvements—exceeding $59$dB—highlighting the effectiveness of the proposed approach.  
\end{abstract}
\ifconf
\else\fi
\section{Introduction}
\label{sec:intro}

Sampling is a cornerstone of signal processing, enabling continuous-time signals to be represented in discrete form and thereby processed digitally \cite{eldar2015sampling,unser2000sampling}. The classical paradigm is the Shannon–Nyquist sampling theorem, which guarantees perfect recovery of bandlimited signals from uniformly spaced samples taken above twice the signal’s bandwidth \cite{nyquist1928certain}. While this theory underlies most modern data converters, its reliance on a global clock and uniform sampling grid often limits efficiency in power- or bandwidth-constrained applications.

\begin{figure}
    \centering
    \includegraphics[width=1\linewidth]{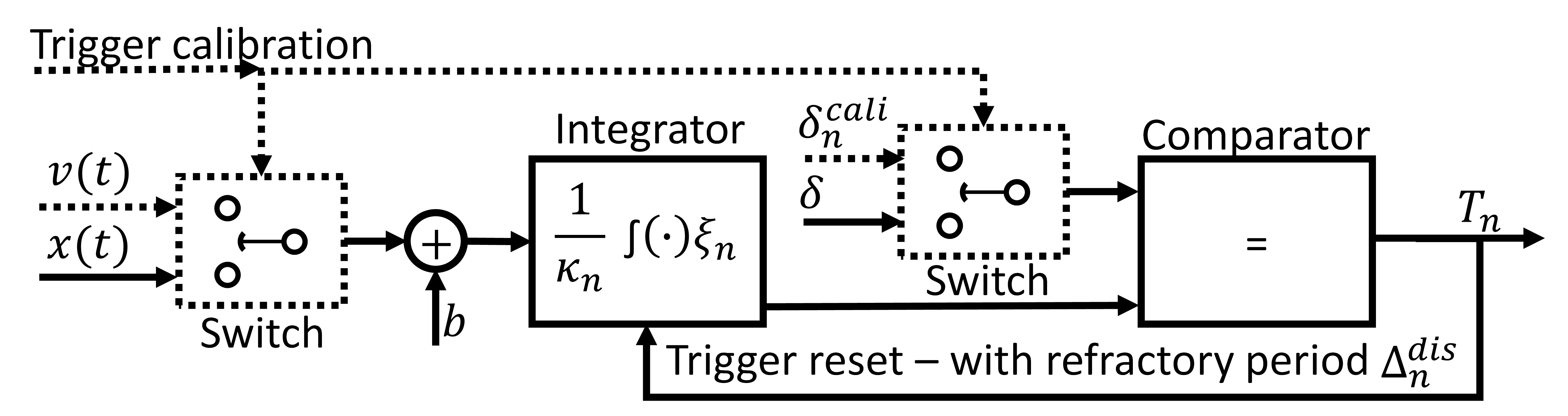}
    \caption{\small IF-TEM design in solid lines under the existing model with known fixed $\kappa_n=\kappa_{n=0}$, $\xi_n=1$ and $\Delta^{\mathrm{dis}}_{n}=\Delta^{\mathrm{dis}}_{n=0}$ \cite{lazar2004time}, and the self-calibrating design, S-IF-TEM in solid with dashed lines under practical model P-IF-TEM with unknown varying $\kappa_n$, $\xi_n$, and $\Delta^{\mathrm{dis}}_{n}$.}
    \label{fig:OffsetEstimation_Scheme}
    \vspace{-0.5cm}
\end{figure}

Time-encoding machines (TEMs), and in particular integrate-and-fire TEMs (IF-TEMs), provide an alternative framework for analog-to-digital conversion (ADC). Instead of recording signal amplitudes at predetermined instants, an IF-TEM encodes information into the timing of events, thereby eliminating the need for a global sampling clock \cite{asynchronous_adc,lazar2004perfect,adam2020sampling,rastogi2009low,alexandru2019reconstructing,hilton2021time,rudresh2020time,adam2020encoding,martinez2019delta,neuromorphic_computer,camera2,naaman_tem,florescu2023generalized,thaoPOCS,kamath2023,liu2023}. As illustrated in Fig~\ref{fig:OffsetEstimation_Scheme}, the basic operation involves biasing the input signal to ensure positivity, integrating the result, and triggering a firing whenever a fixed threshold is reached. Each firing instant is recorded, and the sequence of event times constitutes the digital representation of the signal \cite{lazar2004perfect,lazar2004time}. 

Theoretical advances have established that IF-TEMs can sample and reconstruct bandlimited signals under suitable conditions \cite{lazar2004time,lazar2003time,feichtinger2012approximate,adam2020sampling,adam2019multi,thaoPOCS,liu2023}. In particular, the firing rate is bounded above and below, with the bounds determined by both signal amplitude and system parameters. For perfect recovery, the lower bound on the firing rate must exceed the Nyquist rate of the signal, which often leads to unavoidable oversampling in practice.  The issue of oversampling could be addressed by adaptively changing the bias term \cite{omar2024adaptive, arora2025lowrate}.

A further challenge and a largely unaddressed practical issue is the drift or variation in IF-TEM parameters—such as integration constant \cite{kong2012time,naaman2024hardware}. The drift is due to device mismatches, temperature fluctuations, or long-term variations in the analog circuitry. Such changes directly affect the firing dynamics and, consequently, the fidelity of reconstruction \cite{naaman2024hardware}. This motivates the development of self-calibrating IF-TEMs \cite{kong2012time}, where the system continuously adapts or corrects its parameters to ensure robust and accurate signal encoding.

Such calibration challenges are not unique to IF-TEMs \cite{shi2018evolution,kannadassan2024high,verma2006frequency,tripathi2017mismatch}. In conventional ADCs, mismatch, offset, and gain errors are well-known issues that are routinely handled through self-calibration techniques \cite{moon1997background, thirunakkarasu2014built, chatterjee2022self}, and more recent innovations such as bi-linear homogeneity enforced calibration \cite{wagner2025bi}, and self-calibrating comparators in SAR ADCs \cite{tang2023design}. In the neuromorphic domain, hardware systems also contend with drift and variability—recent work, for instance, uses non-linear current scaling to compensate for drift in phase-change memory elements within spiking neural networks \cite{palhares2024phase}.

In this paper, we introduce a new practical integrate-and-fire time encoding machine (P-IF-TEM) that extends classical system models by explicitly accounting for device mismatches and imperfections that can cause significant reconstruction errors. Unlike the existing IF-TEM model \cite{lazar2004time}, P-IF-TEM considers scenarios where (i) system parameters may be inaccurately known and vary over time (cf. $\kappa_n$ in Fig~\ref{fig:OffsetEstimation_Scheme}), (ii) the integrator discharge  time after firings can vary (cf. $\Delta^{\mathrm{dis}}_{n}$ in Fig~\ref{fig:OffsetEstimation_Scheme}), and (iii) the sampler may operate in its nonlinear region when subject to large input dynamic ranges (cf. $\xi_n$ in Fig~\ref{fig:OffsetEstimation_Scheme} and Sec.~\ref {sec:Problem} for details). Under these practical mismatches, we first establish bounds on the sampling rate and derive conditions that guarantee perfect reconstruction. Building on this analysis, we propose a novel self-calibrating framework as illustrated in Fig~\ref{fig:OffsetEstimation_Scheme} with solid and dashed lines, termed self-calibrating or S-IF-TEM, which operates directly within the practical setting. S-IF-TEM enables simultaneous parameter estimation and signal reconstruction during the sampling process, thereby mitigating the effects of unknown parameters and mismatches (see Sec.~\ref{subsec:estimation_method}). We provide analytical results demonstrating the feasibility of perfect reconstruction in P-IF-TEM under the proposed self-calibration approach (cf. Sec. \ref{sec:theoretical_guarantees}). Specifically, in the proposed self-calibration process, we suggest injecting twice a reference signal $v(t)$ to estimate the unknown parameters and mismatches by selecting the threshold $\delta^{cali}_{n}$ and the injected reference signals $V_n$ that meet the analytical requirements for the maximum possible non-sampling input signal period during the estimation stage. Finally, we provide an evaluation study that validates the significant improvements (in Sec.~\ref{sec:EvaluationResults}). For instance, over 50 synthetic signals tested, the reconstructed signal from the proposed method and input signal differ by an average and maximum normalized mean square error (NMSE) of at most  $-86.52$ dB and $-83.45$ dB, respectively, while the existing IF-TEM approach differs by an average and maximum NMSE of at most $-23.77$ dB and $-18.49$ dB, respectively.


\section{Problem Formulation}\label{sec:Problem}\label{sec:problem_formulation}
Consider a class of finite-energy bandlimited signals $\mathcal{B}_{\omega_M, E}$ such that, for any $x(t) \in \mathcal{B}_{\omega_M, E}$, the Fourier transform of $x(t)$, given as $X(\omega) = \int x(t) \, e^{\mathrm{j}\omega t} \mathrm{d}t$, vanishes outside the interval $[-\omega_M, \omega_M]$. The energy of the signal is bounded as $\int \|x(t)\|^2 \mathrm{d}t \leq E$. Furthermore, the signals in $\mathcal{B}_{\omega_M, E}$ are bounded as $|x(t)|\leq c = \sqrt{{E\, \omega_M}/{\pi}}$ \cite{papoulis1967limits}. Note that any $x(t) \in \mathcal{B}_{\omega_M, E}$ can be perfectly reconstructed from its uniform samples $\{x(nT_s)\}$ \cite{nyquist1928certain,antoniou2006digital}, provided that $T_s < {\pi}/{\omega_M}$. In this work, we consider sampling a signal in $\mathcal{B}_{\omega_M, E}$ using non-ideal IF-TEM methods \cite{lazar2003time, lazar2004perfect}.

In an ideal IF-TEM, the signal to be sampled, $x(t) \in \mathcal{B}_{\omega_M, E}$ with $|x(t)|\leq c$, is first biased to form a non-negative signal $x(t) + b$, where $b>c$. The resulting signal is then integrated to have $\frac{1}{\kappa}\int (x(t)+b) \mathrm{d}t$ where $\kappa>0$ is the integration time constant. The time or firing instant is recorded when the integrator’s output reaches a threshold $\delta$, after which the integrator is reset. This mechanism yields a series of time encodings $\cdots<t_{n-1}<t_n<t_{n+1}<\cdots$, which are related to the signal as
\vspace{-0.2cm}
\begin{align*}
\frac{1}{\kappa} \int_{t_{n}}^{t_{n+1}}\left(x(t)+b\right) \mathrm{d}t = \delta.
\end{align*}
The signal $x(t)$ can be recovered from the time encodings by an iterative algorithm if $T_n = t_{n+1} - t_{n} < {\pi}/{\omega_M}, \forall n$ \cite{lazar2004perfect}.

 In practical IF-TEM implementation, P-IF-TEM, the parameter $\kappa$ may not be known accurately and may vary over time. Further, the integrator is not reset immediately after the firing and may operate in its non-linear region due to a large input dynamic range. To model these variations and non-linearities, we assume that the parameters may change after each firing $t_n$ as discussed in the following.
\begin{enumerate}[wide, labelwidth=0.3cm, labelindent=1pt]
    \item The integration time-constant variation is captured as $\kappa_n = \Kcalc + \KNerr$, where $\Kcalc$ is the time-constant calculated from the resistor and capacitor used in the integrator and $\KNerr$ is the error on the time constant \cite{verma2006frequency,tripathi2017mismatch}. 
    \item The integrator resets after $t_n + \Delta^{\mathrm{dis}}_n$, where $\Delta^{\mathrm{dis}}_n$ is the discharge time of the capacitor after the firing event \cite{verma2006frequency,tripathi2017mismatch}. 
    \item The non-linearity in the integration is difficult to model. Here, we capture it by an unknown function $\xi(t)$ multiplied by the integrand during the integration \cite{kong2012time,sankar2007analysis}. 

    
\end{enumerate} 
In P-IF-TEM, the time encodings are related to $x(t)$ as 
\vspace{-0.1cm}
\begin{align}\label{eq:iftem_practical}
    \frac{1}{\kappa_n} \int_{t_{n-1}+\Delta^{\mathrm{dis}}_n}^{t_n} \xi(t)\left(x(t)+b\right)\ \mathrm{d}t = \delta. 
\end{align}
The operating point of the integrator (linear or non-linear region) depends on the dynamic range of the input $x(t)+b$. To capture this aspect in our model, we rewrite \eqref{eq:iftem_practical} as 
\begin{multline}
 \frac{\gamma_{n}}{\kappa_n} \int_{t_{n}+\Delta^{\mathrm{dis}}_n}^{t_{n+1}}\left(x
(t)+b\right)\, \mathrm{d}t\\
= \frac{1}{\sigma_n} \int_{t_{n}+\Delta^{\mathrm{dis}}_{n}}^{t_{n+1}}\left(x
(t)+b\right)\, \mathrm{d}t =\delta,
\label{eq:iftem_practical2}
\end{multline}
where $\gamma_{n}$ is an averaging constant factor resulting from $\xi(t), t\in[t_{n}+\Delta^{\mathrm{dis}}_n,t_{n+1}]$, and $\sigma_n\triangleq {\kappa_n}/{\gamma_{n}}$ denotes the overall integration period scaling factor that include the mismatches. 

The objective of this work is to determine the unknown parameters $\Delta^{\mathrm{dis}}_n$ and $\sigma_{n}$ at regular intervals of time and correct the values to increase the reconstruction accuracy of the signal from the resulting time encodings. To this end, we assume that the parameters do not change significantly over a fixed and known time period, or in this context, over $l$ consecutive firing instants. With these assumptions, next, we discuss the proposed self-calibration approach.  



\section{Proposed Self-Calibration Approach and Reconstruction Guarantees}
\label{sec:S-IF-TEM}
Here, we discuss the proposed S-IF-TEM approach, followed by the guarantees for perfect reconstruction of the signal from the time encodings of P-IF-TEM and S-IF-TEM.

\subsection{Unknown Parameter Estimation Through Self-Calibration}
\label{subsec:estimation_method}

To calibrate the P-IF-TEM by estimating $\sigma_n$ and $\Delta^{\mathrm{dis}}_{n}$, we intermittently change the input of the TEM from the true signal $x(t)$ to a calibrating signal $v(t)$ as shown in Fig.~\ref{fig:OffsetEstimation_Scheme}. By using the injected signal $v(t)$, we provide herein an approach to estimate ${\sigma}_n$ and ${\Delta}^{\mathrm{dis}}_{n}$. Since the signal's $x(t)$ is not measured during the calibration phase, it has to be measured above the minimum requirements during the non-calibration phase for perfect reconstruction, as elaborated in the following subsection.

\begin{figure}
  \centering
  \includegraphics[width=0.99\linewidth]{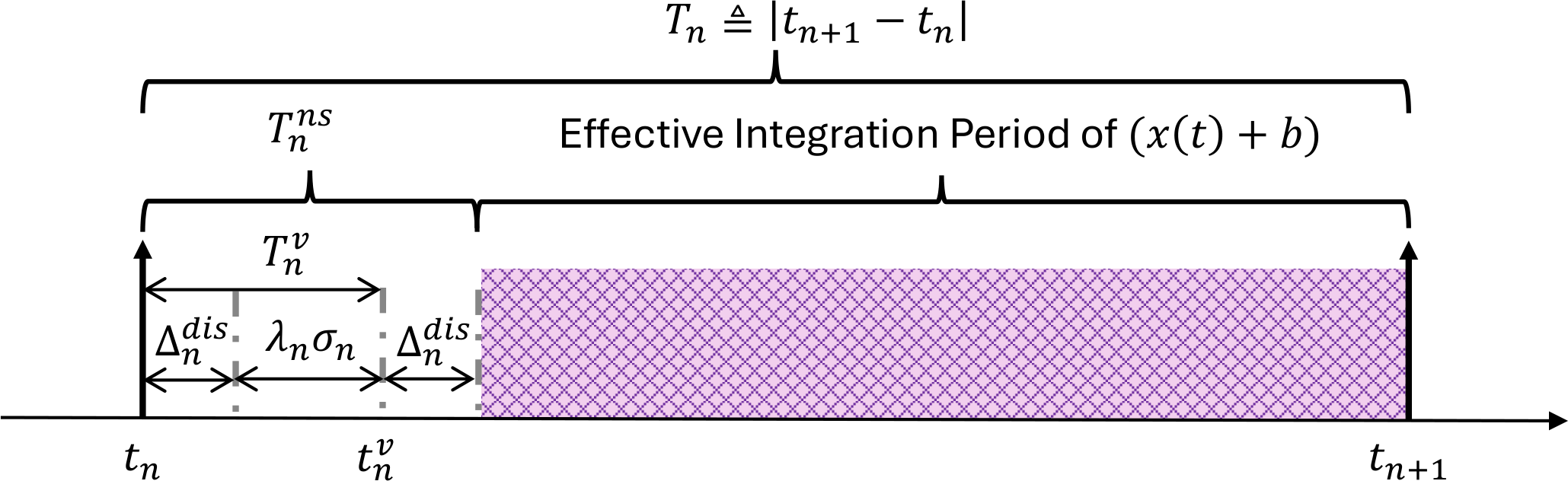}
  \caption{\small Close-up view of a $T_n$ sampling interval illustrating the injection of the reference signal $v(t)$ during $T^v_n$, where $\lambda_n \triangleq\frac{\THcali}{V_n+b}$.}
  \label{fig:v_injection_in_Tn}
  \vspace{-0.5cm}
\end{figure}

For the calibration, we assume that the parameters do not change significantly over the interval $[t_n, t_{n+l+1}]$, and we have
\begin{align}
    {\sigma}_{n} \approx {\sigma}_{n+k+1} \quad \text{and} \quad
     {\Delta}^{\mathrm{dis}}_{n} \approx {\Delta}^{\mathrm{dis}}_{n+k+1}. \label{eq:the_assumption}
\end{align}
To estimate these parameters within these intervals, a calibration voltage $v(t)$ is injected twice at firing instants $t_{n}$ and $t_{ n+k}$, where $k \in \{1, \cdots, l\}$. Specifically, the input to S-IF-TEM is switched from $x(t)$ to $v(t)$ soon after triggers $t_n$ and $t_{n+k}$ such that $v(t_n) = V_n$ and $v(t_{n+k}) = V_{n+k} = \alpha V_n$, where $\alpha \neq 1$. Note that the voltages could be injected after any two firings in the interval considered. During the calibration period, the threshold of the comparator is switched from $\delta$ to $\THcali_n$ and $\THcali_{n+k}$ (cf. Remark~\ref{rem:cond_params} for more details) as shown in Fig~\ref{fig:OffsetEstimation_Scheme}. The signal and threshold would be shifted back to $x(t)$ and $\delta$, respectively, soon after the subsequent firings $t_{n}^v$ and $t_{n+k}^v$. These two firings are a result of the reference voltages $V_n$ and $V_{n+k}$, respectively, which are highlighted by the superscript $v$. By using \eqref{eq:iftem_practical2} with the assumption in \eqref{eq:the_assumption}, for any fixed $n$ in the self-calibration period proposed, the firing intervals 
\begin{align}
\textstyle  T_{n}^v = t_{n}^v - t_n
\, \text{  and  }\, T_{n+k}^v =t_{n+k}^v - t_{n+k}, \label{eq:ref_interval}  
\end{align} depicted in Fig~\ref{fig:v_injection_in_Tn}, are related to the reference voltages and the calibrating thresholds in the sampled stages for estimation as
\begin{align}
\begin{aligned}
\textstyle     T_{n}^v = \frac{\THcali_n}{V_n+b} {\sigma}_{n}+\ {\Delta}^{\mathrm{dis}}_{n}, \text{ and, }
    T_{n+k}^v =
    \frac{\THcali_{n+k}}{\alpha V_{n}+b} {\sigma}_{n}+{\Delta}^{\mathrm{dis}}_{n}.
     \end{aligned} \label{eq:calibration_intervals}
 \end{align}
These are two linear equations with unknowns ${\sigma}_n$ and ${\Delta}^{\mathrm{dis}}_{n}$, and with the rest of the terms known. In the proposed method for self-calibration, the equations are solved to estimate the unknown parameters, and the estimates are given as $\hat{\sigma}_n$ and $\hat{\Delta}^{\mathrm{dis}}_{n}$. In turn, the parameters are used to calibrate the measurements during signal reconstruction \cite{lazar2004perfect,lazar2004time} for the time interval $[t_{n}, t_{n+l+1}]$.

A few remarks are in order. During the intervals in \eqref{eq:ref_interval}, the signal was not measured in addition to the discharge time $\Delta_n^{\mathrm{dis}}$. In \cite{lazar2004time}, the author showed that the signal can be reconstructed from the firings even in the presence of a known constant discharge time, provided that the firing rate is higher than the conventional IF-TEM. We generalize the concept by considering a \emph{non-sampling} time interval $T^{\mathrm{ns}}_n$ after the firing at $t_n$, where the signals are measured during the intervals $[t_n + T^{\mathrm{ns}}_n, t_{n+1}]$. Hence, in S-IF-TEM,  we have that 
\begin{align}
    T^{\mathrm{ns}}_n = 
    \begin{cases}
        \Delta_n^{\mathrm{dis}}, \,\,\, &\text{for the non-calibration phases},\\
        \Delta_n^{\mathrm{dis}} + T^v_{n+p}, \,\,&\text{for the calibration phases},
    \end{cases} \label{eq:Tns}
\end{align}
where $T^v_{n+p}, \, p \in \{0, k\}$. The $\Delta_n^{\mathrm{dis}}$ in the calibration part is due to the $t^v_{n+p}$ firings during the calibration phase (cf. Fig.~\ref{fig:v_injection_in_Tn}). 

As we show in the following section, $T^{\mathrm{ns}}_n$ should be bounded to ensure perfect reconstruction, which in turn requires bounding $T^v_n$ and $T^v_{n+k}$. This will be guaranteed with the appropriate choice of $\THcali_n$ and $\THcali_{n+k}$, as discussed in the next section, specifically, in Remark~\ref{rem:cond_params}. 

\subsection{Theoretical Guarantees for Perfect Reconstruction with Practical Approach and Self-Calibration}
\label{sec:theoretical_guarantees}
In this section, we present the theoretical results for P-IF-TEM and S-IF-TEM with the scaling factors $\sigma_n$ and the discharge periods $\Delta^{\mathrm{dis}}_n$. To derive the guarantees, we consider the following bounds: $\Delta_n^{\mathrm{dis}} \in \bigl[\Delta^{\mathrm{dis},\inf},\, \Delta^{\mathrm{dis},{\sup}}]$, 
$\kappa_n \in \bigl[\kappa^{\inf},\, \kappa^{\sup}\bigr]$, $\gamma_n \in \bigl[\gamma^{\inf},\, \gamma^{\sup}\bigr],$ and $\sigma_n \in \bigl[\sigma^{\inf},\, \sigma^{\sup}\bigr]$, where $\sigma^{\inf} = 
{\kappa^{\inf}}/{\gamma^{\sup}}$ and $\sigma^{\sup} = {\kappa^{\sup}}/{\gamma^{\inf}}.$ These limits on the parameter variations could be determined from the specifications of components used in the circuit implementation and the operating conditions. 
       
 First, we consider the results when $\{\sigma_n, \Delta^{\mathrm{dis}}_n\}$ are known and calibration is not required. These results will set a baseline for the results that require the injection of the calibration signal. The following two results show that with $\{\sigma_n, \Delta^{\mathrm{dis}}_n\}$, firings are possible and they can be bounded. 
       
\begin{theorem}[t-Transform in P-IF-TEM]
\label{thm:t_transform_prop}
For signals in $\mathcal{B}_{\omega_M, E}$, which are bounded by $c$, a P-IF-TEM produces a strictly increasing sequence of spike times 
$\{t_n\}_{n \in \mathbb{Z}}$ that obey the recursive relation
\begin{equation} 
P_n =\int_{t_{n}+\Delta_n^{\mathrm{dis}}}^{\,t_{n+1}} x(s)\,ds
=
\sigma_n \,\delta - b\Big(t_{n+1} - t_{n} - \Delta_n^{\mathrm{dis}}\Big). \label{eq:PIFTEM_averages}
\end{equation}
\end{theorem}
The t-Transform, as introduced in~\cite{lazar2004time}, characterizes how amplitude information is encoded by spike times. In the P-IF-TEM setting, the mapping remains valid, with the encoded information being modulated by the non-ideal integration dynamics  $\xi(t)$. Further, the bounds in P-IF-TEM on the firing intervals are given as follows. 

\begin{proposition}[Inter--Spike Intervals]
\label{cor:Tn_bounds}
For signals in $\mathcal{B}_{\omega_M, E}$, which are bounded by $c$, the intervals $T_n = t_{n+1} - t_{n}$ resulted from P-IF-TEM are bounded as $T_{\min} \leq T \leq T_{\max}$, where 
\begin{align}
\textstyle T_{\min}
     = 
    \frac{\sigma^{\inf}\,\delta}{b + c} + \Delta^{\mathrm{dis}, {\inf}}, \quad T_{\max} =
    \frac{\sigma^{\sup}\,\delta}{b - c} + \Delta^{\mathrm{dis},{\sup}}.  \label{eq:Tn_bounds}
\end{align}
\end{proposition}

The proofs for the previous two results, as well as the following ones, are presented in \cite[Appendix]{SIFTEM2025}.

In a conventional IF-TEM, where the signal is measured in all the time intervals, the upper bound on the firing rate should be less than the Nyquist interval for perfect reconstruction. On the other hand, in P-IF-TEM, while $T_{\max}$ guarantees sufficiently dense sampling as before, $T_{\min}$ restricts the non-sampled portion (cf. \eqref{eq:Tns}). Therefore, the feasibility of reconstruction from the spike times can be reached by accounting for these bounds together, as discussed next. 
\begin{theorem}[Recovery Condition for P-IF-TEM]
\label{thm:rec_condition}
Any signal in $ \mathcal{B}_{\omega_M, E}$ bounded by $c$ can be perfectly recovered from its P-IF-TEM's time-encodings (as in \eqref{eq:PIFTEM_averages}) if $ r + \varepsilon\,(1+r) <1$,
where, $r = {T_{\max}}/{T_{\mathrm{nyq}}}$, $\varepsilon = \sqrt{{\Delta^{\mathrm{dis}, {\sup}}}/{T_{\min}}}$, $T_{\mathrm{nyq}} = {\pi}/{\omega_m}$, and, $T_{\min}$ and $T_{\max}$ are as given in \eqref{eq:Tn_bounds}.
\end{theorem}
The aforementioned result relies on the known bounds of $\{\sigma_n, \Delta^{\mathrm{dis}}_n\}$ and does not consider the calibration process as discussed in Section~\ref{subsec:estimation_method}. With calibration, the non-sampling durations are as given in \eqref{eq:Tns} and illustrated in Fig~\ref{fig:v_injection_in_Tn}, which may differ from $\Delta^{\mathrm{dis}}_n$. Hence, the measurements $P_n$, the spike interval bounds in \eqref{eq:Tn_bounds}, and the reconstruction conditions stated in theorem~\ref{thm:rec_condition} would change. Specifically, the measurements and the firing bounds are given as follows
\begin{equation} 
P_n^c=\int_{t_{n}+T_n^{\mathrm{ns}}}^{\,t_{n+1}} x(s)\,ds
=
\sigma_n \,\delta - b\Big(t_{n+1} - t_{n} - T^{\mathrm{ns}}_n\Big), \label{eq:PIFTEM_averages_cali}
\end{equation}
\vspace{-0.2cm}
\begin{align}
\textstyle T_{\min}^c
     = 
    \frac{\sigma^{\inf}\,\delta}{b + c} + T^{\mathrm{ns},\inf}, \quad T_{\max}^c =
    \frac{\sigma^{\sup}\,\delta}{b - c} + T^{\mathrm{ns},\sup},  \label{eq:Tn_bounds_cali}
\end{align}
where the superscript $c$ highlight the calibration. The main results are summarized in the following theorem.
\begin{theorem}[S-IF-TEM Mechanism and Reconstruction] \label{thm:PIFTEM_calibration}
   Consider a S-IF-TEM with a calibration process and non-sampling intervals bounded as $T^{\mathrm{ns}}_n \in [T^{\mathrm{ns},\inf}, T^{\mathrm{ns},\sup}]$. Then a signal $x(t) \in \mathcal{B}_{\omega_m, E}$, can be reconstructed perfectly from the firings by using the measurements in \eqref{eq:PIFTEM_averages_cali} if
   \[
\textstyle   r^c + \varepsilon^c (1+r^c) <1,
   \]
   where $r^c = {T_{\max}^c}/{T_{\mathrm{nyq}}}$, $\varepsilon^c = \sqrt{{T^{\mathrm{ns},\sup}}/{T_{\min}^c}}$, $T_{\mathrm{nyq}} = {\pi}/{\omega_m}$, and, $T^{c}_{\min}$ and $T^{c}_{\max}$ are as given in \eqref{eq:Tn_bounds_cali}.
\end{theorem} 
 Note that $T^{\mathrm{ns},\inf} = \Delta^{\mathrm{dis},\inf}$ is fixed by the known circuit parameters, while 
$T^{\mathrm{ns},\sup} \triangleq \max\{T_n^{\mathrm{ns}} : r^c + \varepsilon^c(1+r^c) < 1\}$ has design freedom through sampler parameters that apply when sampling $x(t)$.
 \off{Note that $T^{\mathrm{ns},\inf}  = \Delta^{\mathrm{dis}, \inf}$, which is known and fixed for a given circuit realization and operating conditions. On the other hand, $T^{\mathrm{ns},\sup}$ is given as 
\begin{align}
 T^{\mathrm{ns},\sup} = \Delta^{\mathrm{dis}, \sup} + T^{v, \sup}, 
\end{align}
where $T^{v, \sup}$ is derived from \eqref{eq:calibration_intervals}
\begin{align}
\label{eq:T^v_bound}
    T^{v, \sup} = \frac{\delta^{cali, \sup}}{V^{\min} + b}\sigma^{\sup} + \Delta^{\mathrm{dis}, \sup},
\end{align}
where $\delta^{cali, \sup}$ and $V^{\min}$ are the maximum calibration threshold and minimum calibration signal used, respectively, during the calibration phase.} 
\begin{proof} 
The result follows directly by substituting $T_{n}^{\mathrm{ns}}$ in place of $\Delta_{n}^{\mathrm{dis}}$.  
We evaluate $r^{c}$ and $\varepsilon^{c}$ within Theorem~\ref{thm:rec_condition}, thereby ensuring that the reconstruction conditions remain satisfied.  
Moreover, applying  $T_{n}^{\mathrm{ns}}$ to Theorem~\ref{thm:t_transform_prop} yields the corresponding adjusted $t$--transform.
\end{proof}
\begin{remark}\label{rem:cond_params}
Within the calibration process of Section~\ref{subsec:estimation_method}, perfect reconstruction requires the parameter selection as follows. For a given segment, the calibration voltages $V_{n+p}$, $p\in \{0,k\}$, are selected to remain within the same operating region as $x(t)$ in order to capture the actual nonlinear effects. Together with the corresponding thresholds $\delta^{cali}_{n+p}$, $p\in \{0,k\}$, these voltages are selected so that the reconstruction conditions of Theorem~\ref{thm:PIFTEM_calibration} are satisfied. Accordingly, in S-IF-TEM, the calibration parameters are further constrained so that the total calibration duration, given by $T^{v,\sup}_{n+p} + \Delta^{\mathrm{dis},\sup}$,$p\in \{0,k\}$, remains strictly below the maximum admissible non-sampling interval $T^{\mathrm{ns},\sup}$, i.e., $
T^{v,\sup}_{n+p} + \Delta^{\mathrm{dis},\sup} < T^{\mathrm{ns},\sup}$,
where $T^{v,\sup}_{n+p} = \sigma^{\sup}\,{\delta^{cali}_{n+p}}/{(V_{n+p}+b)} + \Delta^{\mathrm{dis},\sup}$ for $p \in \{0,k\}$. Hence, the choice of $V_{n+p}$ and $\delta^{cali}_{n+p}$ must satisfy
\begin{align*}
\label{eq:calibration_bound_sigma}
\textstyle    \frac{\delta^{cali}_{n+p}}{V_{n+p} + b}
    \;<\;
    \frac{T^{\mathrm{ns},\sup} - 2\Delta^{\mathrm{dis}, \sup}}{\sigma^{\sup}}.
\end{align*}
\end{remark}
Following the restrictions given after Theorem~\ref{thm:PIFTEM_calibration}, the proposed method, as described in Section~\ref{subsec:estimation_method}, guarantees perfect reconstruction. 
Incorporating the self-calibration intervals in \eqref{eq:ref_interval}, the measurements and estimations are then used by an iterative algorithm \cite{lazar2004time,lazar2004perfect} to reconstruct the signal. 

\section{Evaluation Results}\label{sec:EvaluationResults}
The performance of the S-IF-TEM sampler was evaluated using synthetic MATLAB signals.
The signals were generated as a finite sum of shifted sinc pulses with random coefficients $c_m$ and spacing of
$T_{nyq} = {2\pi}/{\omega_M}$, given by
\begin{equation*}
\label{eq: sim_signal}
\textstyle    x(t) = \sum_{m=0}^{2M} c_m \ \mathrm{sinc}\!\left(\frac{t - mT_{nyq}}{T_{nyq}}\right).
\end{equation*} 
We evaluate the signals reconstruction by comparing four samplers, with $P_n$ and $P_n^c$ defined in~\eqref{eq:PIFTEM_averages} and~\eqref{eq:PIFTEM_averages_cali}, respectively: 
\begin{itemize}[wide, labelwidth=0.1cm, labelindent=1pt]
    \item \textbf{Ideal IF-TEM:} 
    P-IF-TEM model with $\xi_n$, $\kappa_n$, and $\Delta_n^{\mathrm{dis}}$ known and applied accurately to $P_n$.
    \item \textbf{Blind IF-TEM:} 
    P-IF-TEM model with $\kappa_n$ and $\Delta_n^{\mathrm{dis}}$ fixed at $\kappa_{\text{calc}}$ and $\Delta^{\mathrm{dis}}_{0}$~\cite{lazar2004time}, integrator non-linearity ignored (i.e., fixed $\xi_n=1$), and $P_n$ computed from non-accurate values.
    \item \textbf{S-IF-TEM:} 
    S-IF-TEM proposed herein preforming calibration as in Section~\ref{sec:S-IF-TEM}, using estimated $\hat{\sigma}_n$ and $\hat{T}_n^{\mathrm{ns}}$ in $P_n^{c}$.
    \item \textbf{Genie S-IF-TEM:} 
    S-IF-TEM model, with calibration as in Section~\ref{sec:S-IF-TEM}, but with true ${\sigma}_n$ and ${T}_n^{\mathrm{ns}}$ applied to $P_n^{c}$, serving as an upper bound for comparison on the proposed estimator.
\end{itemize}

\begin{figure}
  \centering
  \includegraphics[width=1\linewidth]{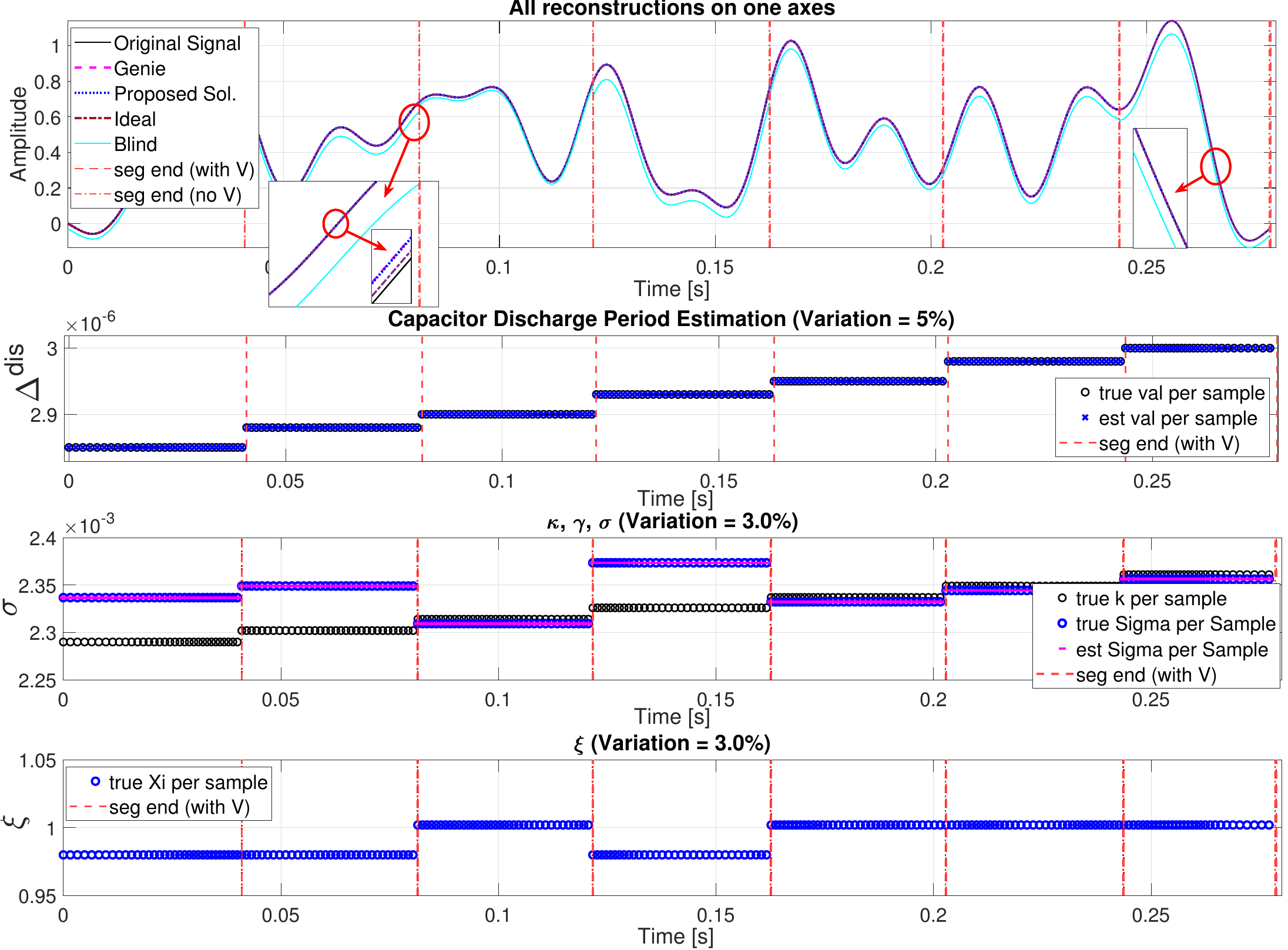}%
  \caption{\small Performance comparison with input signal at 50 Hz.}
  
  \label{fig:50Hz}
  \vspace{-0.5cm}
\end{figure}


In the following results, synthetic signals are bandlimited to $\omega_M = 200\pi$~rad/s (equivalently, $f_{\max}=50$~Hz) and generated with $M=12$. In all cases, the sampler threshold is fixed at $\delta=1$, while $c=\max(|x(t)|)$ is individually determined to capture the amplitude range of each signal. The bias is then proportionally assigned to $b=1.3\cdot c$. Further, we set $k=2$ for the S-IF-TEM and Genie S-IF-TEM samplers. We account for variations in $\xi_n$, $\kappa_n$, and $\Delta^{\mathrm{dis}}_n$,
treating them as fixed within each $40$~ms interval. Note: the fixed values may vary both across signals and between corresponding segments. Specifically, $\Delta_n^{\mathrm{dis}} \in [2.85,\,3]~\mu\text{s}$ (a $5\%$ variation), and the gain parameter takes discrete values $\xi_n \in \{0.98,\,1,\,1.002\}$ depending on the signal amplitude. The integration time-constant $\kappa_n$ lies in $[\kappa^{\inf},\,\kappa^{\sup}]$, with $\kappa^{\inf}$ set to ensure a $3\%$ range. For each signal, $\kappa^{\sup}$ is chosen so that $r+\epsilon(1+r)\approx0.8$, enabling suitable NMSE comparison across signals.


Testing with 50 synthetic signals shows that the proposed S-IF-TEM method and Genie IF-TEM deviate from the original input by at most $-86.52~\mathrm{dB}$ and $-86.48~\mathrm{dB}$ in average NMSE, and $-83.45~\mathrm{dB}$ and $-82.92~\mathrm{dB}$ in maximum NMSE, respectively. The Ideal IF-TEM exhibits slightly better performance, with deviations up to $-86.91~\mathrm{dB}$ on average and $-83.40~\mathrm{dB}$ at worst. This corresponds to an average performance gap of $1.75~\mathrm{dB}$ and a maximum gap of $4.25~\mathrm{dB}$ between the proposed method and the Ideal case. This is unlike the traditional Blind IF-TEM technique, which significantly deviates from the original input by as much as $-18.49~\mathrm{dB}$ on average and $-23.77~\mathrm{dB}$ in the best case.

Fig.~\ref{fig:50Hz} presents an example of a single synthetic signal, particularly with $b = 1.48$, $T^{\mathrm{ns},\sup} = 26~\mu\text{s}$, and $\kappa_n \in (0.00229,\,0.002361)$. The plots show: (a) the reconstruction results for the four IF-TEM samplers compared with the original signal, (b) the true and estimated $\Delta^{\mathrm{dis}}_n$, (c) the true $\kappa_n$ and the true and estimated $\sigma_n$, and (d) the true $\xi_n$.



The results highlight the need for regular calibration of the IF-TEM to mitigate the time-varying nature of the parameters and the effectiveness of the proposed S-IF-TEM.

\section{Conclusions}
We presented a self-calibrating integrate-and-fire time encoding machine, S-IF-TEM, that addresses the limitations of classical existing IF-TEMs under parameter drift and circuit non-idealities. By introducing a practical IF-TEM model, P-IF-TEM, and establishing its theoretical guarantees, we designed a calibration framework that simultaneously estimates parameters and reconstructs the input signal using reference injections. Analysis and simulations confirm that the proposed approach significantly improves reconstruction accuracy, highlighting self-calibration as a key enabler for robust and practical time-based sampling systems.

\ifconf
    \bibliographystyle{IEEEbib}
\else
    \bibliographystyle{IEEEtran}
\fi
\bibliography{strings,refs}


\newpage

\newpage 
\appendix
\ifconf
\begin{center}
    \normalsize\bfseries APPENDIX
\end{center}
\else\fi
\subsection{Definitions}

\begin{definition}
\label{def:timing_def}
The P-IF-TEM output is a strictly increasing sequence of spike times 
$\{t_n \mid n=0,1,\dots,N\}$ over a finite observation window $W = [t_0,t_N]$, 
where $t_0$ denotes the sampling start time. The inter-spike intervals are 
$T_n = t_{n+1} - t_{n}$ and each sample is associated with the interval 
$t \in [\,t_{n} + \Delta^{\mathrm{dis}}_n,\, t_{n+1}\,]$.
\end{definition}
For perfect reconstruction proofs given in Appendix~\ref{sec:app_proofs}, we assume $N\rightarrow\infty$. For a fine regime with segments, one can use the techniques given in \cite{omar2024adaptive}. We keep this interesting direction for future study.

\begin{definition}
\label{def:A_operator}
The operator $\mathcal{A}$ and its adjoint $\mathcal{A}^*$ are defined as

\begin{equation*} \begin{aligned} \mathcal{A}x &= \sum_{n=1}^N P_n \, g\!\left(t - \theta_n\right), & \mathcal{A}^*x &= \sum_{n=1}^N x(\theta_n)\, \mathcal{P}1_{[t_{n}+\Delta_{n}^{dis}, t_{n+1}]}, \end{aligned} \end{equation*}

where $P_n$ is as given in~\eqref{eq:PIFTEM_averages}, 
$g(t) = {\sin(\Omega t)}/{\pi t}$ is the sinc function, 
$\theta_n = \tfrac{1}{2}(t_{n}+t_{n+1})$, 
and where the projection operator $\mathcal{P}$ is given by 
$\mathcal{P}\,\mathbf{1}_{[t_{n}+\Delta_n^{\mathrm{dis}},\,t_{n+1}]}(t) 
= \big(g \ast \mathbf{1}_{[t_{n}+\Delta_n^{\mathrm{dis}},\,t_{n+1}]}\big)(t),$
where $\ast$ denotes convolution and 
$\mathbf{1}_{[t_{n}+\Delta_n^{\mathrm{dis}},\,t_{n+1}]}$ is the unit pulse on 
$[t_{n}+\Delta_n^{\mathrm{dis}},\,t_{n+1}]$.

\end{definition}

\subsection{Proofs of P-IF-TEM Main Results} \label{sec:app_proofs}


\begin{proof}[Proof of Theorem~\ref{thm:t_transform_prop}]
The result follows directly from~\eqref{eq:iftem_practical2} by multiplying both sides by $\sigma_n$ and separating the integral into signal and bias components. Hence, we obtain
\begin{equation*}
\int_{t_{n}+\Delta_n^{\mathrm{dis}}}^{t_{n+1}} x(t)\,dt
+ b\Big(t_{n+1} - t_{n} - \Delta_n^{\mathrm{dis}}\Big)
= \sigma_n \delta.
\end{equation*}
Rearranging, we obtain
\begin{equation*}
P_n \;\triangleq\;\int_{t_{n}+\Delta_n^{\mathrm{dis}}}^{t_{n+1}} x(t)\,dt
= \sigma_n \delta - b\Big(t_{n+1} - t_{n} - \Delta_n^{\mathrm{dis}}\Big),
\end{equation*}
which completes the proof.
\end{proof}

\begin{proof}[Proof of Proposition~\ref{cor:Tn_bounds}]
For each sample $n$, and using $|x(t)| \leq c_{\max}$,  
the integral in~\eqref{eq:PIFTEM_averages} can be bounded as
\begin{multline*}
-\,c_{\max}\,\big(t_{n+1}-t_{n}-\Delta_n^{\mathrm{dis}}\big)
\;\le\; \int_{t_{n}+\Delta_n^{\mathrm{dis}}}^{t_{n+1}} x(t)\,dt
\\ \;\le\; c_{\max}\,\big(t_{n+1}-t_{n}-\Delta_n^{\mathrm{dis}}\big).
\end{multline*}
Substituting the integral from~\eqref{eq:PIFTEM_averages}, gives
\begin{multline*}
-\,c_{\max}\,\big(t_{n+1}-t_{n}-\Delta_n^{\mathrm{dis}}\big)
\le\frac{\kappa_n}{\gamma_n}\,\delta 
   - b\big(t_{n+1} - t_{n} - \Delta_n^{\mathrm{dis}}\big) \\
\le c_{\max}\,\big(t_{n+1}-t_{n}-\Delta_n^{\mathrm{dis}}\big).
\end{multline*}
Rearranging and isolating $T_n = t_{n+1} - t_{n}$ yields the bounds
\begin{multline*}
T_{\min}
\;\triangleq\; 
\frac{\kappa_{n}\,\delta}{\gamma_n (b + c_{\max})} 
+ \Delta_{n}^{\mathrm{dis}}
\;\leq\; T_n \\[0.5em]
\;\leq\;
\frac{\kappa_{n}\,\delta}{\gamma_n (b - c_{\max})} 
+ \Delta_{n}^{\mathrm{dis}}
\;\triangleq\;
T_{\max}.
\end{multline*}

Now, considering settings where the parameter ranges are known, as described in in ~\ref{sec:theoretical_guarantees}, we obtain
\begin{multline*}
T_{\min}
     \;\triangleq\; 
    \frac{\kappa^{\inf}\,\delta}{\gamma^{\sup}\,\bigl(b + c_{\max}\bigr)} 
    \;+\; \Delta^{\mathrm{dis},\inf}
    \;\leq\; T_n \\[0.5em]
    \;\leq\;
    \frac{\kappa^{\sup}\,\delta}{\gamma^{\inf}\,\bigl(b - c_{\max}\bigr)} 
    \;+\; \Delta^{\mathrm{dis},\sup}
    \;\triangleq\; T_{\max}.
\end{multline*}
Finally, substituting $\sigma^{\inf} \;\triangleq\; {\kappa^{\inf}}/{\gamma^{\sup}}$ and
$\sigma^{\sup} \;\triangleq\; {\kappa^{\sup}}/{\gamma^{\inf}}$,
we obtain the bounds on $T_{\min}$ and $ T_{\max}$ in~\eqref{eq:Tn_bounds}, which completes the proof.
\end{proof}

\begin{proof}[Proof of Theorem~\ref{thm:rec_condition}]
Following the same techniques in~\cite[Proposition 1]{lazar2004time}, we generalize the reconstruction condition for P-IF-TEM. 
Let $\mathcal{A}$ and its adjoint $\mathcal{A}^*$ be as defined in Definition~\ref{def:A_operator}, 
under the sampling framework specified in Definition~\ref{def:timing_def}. 
The squared windowed $L^2$-norm is defined as
\[
\|f\|_{W}^2 \triangleq \int_{t_{\mathrm{start}}}^{t_{\mathrm{end}}} |f(u)|^2\,du,
\]
with $W=[t_0,t_N]$ in our settings, corresponding to $N$ samples. The signal $x(t)$ can be iteratively reconstructed via the Neumann series of $\mathcal{A}$ 
if the following condition holds,
$|I - \mathcal{A}\|_{W} < 1$.

We apply the adjoint operator $\mathcal{A}^*$ to $x$ and bound
\begin{align*}
&\bigl\| x - \mathcal{A}^* x \bigr\|_{W} = \left\|\, x - \sum_{n=1}^N x(\theta_n)\,\mathcal{P}\,1_{[t_{n}+\Delta_n^{\mathrm{dis}},\,t_{n+1})} \,\right\|_{W} \\
&{\leq} \left\|\, x - \sum_{n=1}^N x(\theta_n)\,1_{[t_{n}+\Delta_n^{\mathrm{dis}},\,t_{n+1})} \,\right\|_{W} \\
&= \Bigg\| \sum_{n=1}^N \bigl(x - x(\theta_n)\bigr)\,1_{[t_{n},\,t_{n+1})}\\ 
& \hspace{1.5cm} + \sum_{n=1}^N x(\theta_n)\,1_{[t_{n},\,t_{n}+\Delta_n^{\mathrm{dis}})} \Bigg\|_{W} \\
&\le \left\| \sum_{n=1}^N \bigl(x - x(\theta_n)\bigr)\,1_{[t_{n},\,t_{n+1})}       \right\|_{W} \\
& \hspace{1.5cm}  + \left\| \sum_{n=1}^N x(\theta_n)\,1_{[t_{n},\,t_{n} + \Delta_n^{\mathrm{dis}})} \right\|_{W}.
\end{align*}
From the decomposition above, the first term can be bounded as
\[
\left\| \sum_{n=1}^N \bigl(x - x(\theta_n)\bigr)\,1_{[t_{n},\,t_{n+1})} \right\|_{W} 
\;\leq\; r \|x\|_{W},
\]
while the second term satisfies
\[
\left\| \sum_{n=1}^N x(\theta_n)\,1_{[t_{n},\,t_{n}+\Delta_n^{\mathrm{dis}})} \right\|_{W} \leq \varepsilon\left\| \sum_{n=1}^N x(\theta_n)\,1_{[t_{n},\,t_{n+1})} \right\|_{W}.
\]
Now, noting that 
\begin{align*}
&\left\|\sum_{n=1}^N x(\theta_n)\,1_{[t_{n},\,t_{n+1})}\right\|_{W}\\
& = \;\; \left\| x - \sum_{n=1}^N \bigl(x - x(\theta_n)\bigr)\,1_{[t_{n},\,t_{n+1})} \right\|_{W} \\
&\le \|x\|_{W} \;+\; \left\| \sum_{n=1}^N \bigl(x - x(\theta_n)\bigr)\,1_{[t_{n},\,t_{n+1})} \right\|_{W} \\
&\le (1+r)\,\|x\|_{W}.
\end{align*}
Therefore, $\mathcal{A}$ satisfies
\begin{align*}
\bigl\| x - \mathcal{A} x \bigr\|_{W} 
&\le \; r \|x\|_{W} \;+\; \varepsilon \bigl(1 + r\bigr)\,\|x\|_{W},
\end{align*}
such that a sufficient condition for reconstruction is given by
\[
r+\varepsilon\,(1+r) \;<\; 1.
\]
\end{proof}
\end{document}